# The Generalized Uncertainty Principle and Black Hole Remnants*

Ronald J. Adler

Gravity Probe B, W. W. Hansen Experimental Physics Laboratory
Stanford University, Stanford CA 94035

Pisin Chen

Stanford Linear Accelerator Center
Stanford University, Stanford CA 94309

David I. Santiago

Department of Physics
Stanford University, Stanford CA 94035

## Summary

In the current standard viewpoint small black holes are believed to emit black body radiation at the Hawking temperature, at least until they approach Planck size, after which their fate is open to conjecture. A cogent argument against the existence of remnants is that, since no evident quantum number prevents it, black holes should radiate completely away to photons and other ordinary stable particles and vacuum, like any unstable quantum system. Here we argue the contrary, that the generalized uncertainty principle may prevent their total evaporation in exactly the same way that the uncertainty principle prevents the hydrogen atom from total collapse: the collapse is prevented, not by symmetry, but by dynamics, as a minimum size and mass are approached.

# The Generalized Uncertainty Principle and Black Hole Remnants

In the standard view of black hole thermodynamics, based on the entropy expression of Bekenstein [1] and the temperature expression of Hawking [2], a small black hole should emit black body radiation, thereby becoming lighter and hotter, and so on, leading to an explosive end when the mass approaches zero. However Hawking's calculation assumes a classical background metric for the black hole and ignores the energy of the radiation compared to the rest energy of the black hole, assumptions which must break down as the black hole becomes very small and light. Thus it does not provide an answer as to whether a small black hole should evaporate entirely to photons and other ordinary particles and vacuum, or leave something else behind, which we refer to as a remnant.

Numerous calculations of black hole radiation properties have been made from different points of view [3], and some hint at the existence of remnants, but in the absence of a well-defined quantum gravity theory none appears to give a definitive answer.

A cogent argument against the existence of remnants can be made [4]: since there is no evident symmetry or quantum number preventing it, a black hole should radiate entirely away to photons and other ordinary stable particles and vacuum, just like any unstable quantum system.

We here argue the contrary, that the total collapse of a black hole may be prevented by dynamics, and not by symmetry. Just as we may consider the hydrogen atom to be prevented from collapse by the uncertainty principle [5] we argue that the generalized uncertainty principle (GUP) may prevent a black hole from complete evaporation.

The uncertainty principle argument for the stability of the hydrogen atom can be stated very briefly. The energy of the electron is $p^2/2m - e^2/r$, so the classical minimum energy is very large and negative, corresponding to the configuration $p = r = 0$, which is not compatible with the uncertainty principle. If we impose as a minimum condition that $p \approx \hbar/r$ we see that

$$E = \frac{\hbar^2}{2mr^2} - \frac{e^2}{r}, \quad \text{thus} \quad r_{\min} = \frac{\hbar^2}{me^2}, \quad E_{\min} = -\frac{me^4}{2\hbar^2}. \tag{1}$$

That is the energy has a minimum, the correct Rydberg energy, when $r$ is the Bohr radius, so the atom is stabilized by the uncertainty principle.

The GUP gives the position uncertainty as

$$\Delta x \geq \frac{h}{\Delta p} + L_p^2 \frac{\Delta p}{h} \quad , \quad L_p = \sqrt{\frac{Gh}{c^3}} \quad , \quad \text{(Planck distance)} \, . \qquad (2)$$

This is a result of string theory [6] or more general considerations of quantum mechanics and gravity [7]. A heuristic derivation may also be made on dimensional grounds. We think of a particle such as an electron being observed by means of a photon of momentum $p$. The usual Heisenberg argument leads to an electron position uncertainty given by the first term of (2). But we should add to this a term due to the gravitational interaction of the electron with the photon, and that term must be proportional to $G$ times the photon energy, or $Gpc$. Since the electron momentum uncertainty $\Delta p$ will be of order of $p$, we see that on dimensional grounds the extra term must be of order $G\Delta p / c^3$, as given in (2). Note that there is no h in the extra term when expressed in this way. The position uncertainty has a minimum value of $\Delta x = 2L_p$, so the Planck distance plays the role of a minimum or fundamental distance.

The Hawking temperature for a spherically symmetric black hole may be obtained in a heuristic way with the use of the standard uncertainty principle and general properties of black holes [8]. We picture the quantum vacuum as a fluctuating sea of virtual particles; the virtual particles cannot normally be directly observed without violating energy conservation. But near the surface of a black hole the effective potential energy can negate the rest energy of a particle and give it zero total energy, and the surface itself is a one-way membrane which can swallow particles so that they are henceforth not observable from outside. The net effect is that for a pair of photons one photon may be absorbed by the black hole with effective negative energy $-E$, and the other may be emitted to asymptotic distances with positive energy $+E$. The characteristic energy $E$ of the emitted photons may be estimated from the standard uncertainty principle. In the vicinity of the black hole surface there is an intrinsic uncertainty in the position of any particle of about the Schwarzschild radius, $r_s$, due to the behavior of its field lines [9], as well as on dimensional grounds. This leads to momentum uncertainty

$$\Delta p \approx \frac{h}{\Delta x} = \frac{h}{2r_s} = \frac{hc^2}{4GM} \quad , \quad \Delta x = r_s = \frac{2GM}{c^2} \quad , \qquad (3)$$

and to an energy uncertainty of $\Delta pc = \hbar c^3 / 4GM$. We identify this as the characteristic energy of the emitted photon, and thus as a characteristic temperature; it agrees with the Hawking temperature up to a factor of $2\pi$, which we will henceforth include as a "calibration factor" and write, with $k_B = 1$,

$$T_H \approx \frac{\hbar c^3}{8\pi GM} = \frac{M_P^2 c^2}{8\pi M} \quad , \qquad M_P = \sqrt{\frac{\hbar c}{G}} \quad . \tag{4}$$

We know of no way to show heuristically that the emitted photons should have a thermal black body spectrum except on the basis of thermodynamic consistency.

We may use the GUP to derive a modified black hole temperature exactly as above. From (2) we solve for the momentum uncertainty in terms of the distance uncertainty, which we again take to be the Schwarzschild radius $r_s$. This gives the following momentum and temperature for radiated photons

$$\frac{\Delta p}{\hbar} = \frac{\Delta x}{2L_P^2}\left[1 \mp \sqrt{1 - 4L_P^2/\Delta x^2}\right] \quad , \qquad T_{GUP} = \frac{Mc^2}{4\pi}\left[1 \mp \sqrt{1 - M_P^2/M^2}\right] \quad , \tag{5}$$

where we have again inserted the "calibration factor" of $2\pi$. This agrees with the standard result (4) for large mass if the negative sign is chosen, whereas the positive sign has no evident physical meaning. However the temperature becomes complex and unphysical for mass less than the Planck mass and Schwarzschild radius than $2L_P$, the minimum size allowed by the GUP. At the Planck mass the slope is infinite, corresponding to zero heat capacity of the black hole. The temperature as a function of mass is shown in fig. 1.

The entropy is obtained by integration of $dS = c^2 TdM$, and we obtain the standard Bekenstein entropy and a modified GUP entropy from (4) and (5) respectively,

$$S_B = \frac{4\pi GM^2}{\hbar c} = 4\pi \frac{M^2}{M_P^2} \quad , \tag{6a}$$

$$S_{GUP} = 2\pi\left[\frac{M^2}{M_P^2}\left(1 - \frac{M_P^2}{M^2} + \sqrt{1 - \frac{M_P^2}{M^2}}\right) - \log\left(\frac{M + \sqrt{M^2 - M_P^2}}{M_P}\right)\right] \quad . \tag{6b}$$

We have normalized the modified entropy to zero at $M_P$, as shown in fig. 2.

A black hole whose temperature is greater than the ambient temperature, about 2.7K for the present universe, should radiate energy in the form of photons and other ordinary particles, thereby reducing its mass further and increasing its temperature. If we assume the energy loss is dominated by photons we may use the Stefan-Boltzmann law to estimate the mass and energy output as functions of time. For the standard case this leads to

$$\frac{d}{dt}\left(\frac{M}{M_P}\right) = -\left(\frac{M_P}{M}\right)^2 \frac{1}{60(16)^2 \pi T_P}, \quad \text{or} \quad \frac{dx}{dt} = -\frac{1}{x^2 t_{ch}},$$

$$x = \left[x_i^3 - \frac{3t}{t_{ch}}\right]^{1/3}, \quad \frac{dx}{dt} = \frac{-1}{t_{ch}\left(x_i^3 - 3t/t_{ch}\right)^{2/3}}, \quad \text{(standard case)}, \quad (7a)$$

where $x = M/M_P$ and the characteristic time is $t_{ch} = 60(16)^4 \pi T_P$, which is about $4.8 \times 10^4$ times the Planck time, $T_P = \sqrt{hG/c^5}$; here $x_i$ refers to the initial mass of the hole. The black hole evaporates to zero mass in time $t/t_{ch} = (M_i/M_P)^3/3$, and the energy radiated has an infinite spike at the end of the process. For the modified case we obtain

$$\frac{dx}{dt} = -\frac{16x^6}{t_{ch}}\left(1 - \sqrt{1 - \frac{1}{x^2}}\right)^4,$$

$$\frac{t}{t_{ch}} = \frac{1}{16}\left[\frac{8}{3}x^3 - 8x - \frac{1}{x} + \frac{8}{3}(x^2-1)^{3/2} - 4\sqrt{x^2-1} + 4\arccos\frac{1}{x}\right]_{M/m_p}^{M_i/m_p} \quad \text{(GUP case)}.$$

(7b)

The masses and energy outputs given by (7a) and (7b) are shown in figures 3 and 4; in the modified case the output is finite at the end point when $x = 1$ and is given by $dx/dt = -16/t_{ch}$, whereas for the standard case it is infinite at the endpoint when $x = 0$. The modified results thus appear to be more physically reasonable than the standard results.

The picture that follows from the above results is that a small black hole, with temperature greater than the ambient temperature, should radiate photons, as well as other ordinary particles, until it approaches Planck mass and size. At the Planck scale it ceases to radiate and its entropy reaches zero, even though its effective temperature reaches a maximum. It cannot radiate further and becomes an inert remnant, possessing only gravitational interactions. Note that, as pointed out by York [3], the remnants need not have a classical black hole horizon structure. Such remnants may have been in existence since very early in the history of the universe and are an attractive dark matter candidate [10].

As with other calculations dealing with Hawking radiation we have not treated all of the gravitational aspects of the problem completely consistently. That is we have not taken account of the recoil of the black hole when radiating very high energy particles, possible quantization of the black hole mass and metric, etc. etc. [11]. Thus, while we cannot expect our results to incorporate all aspects of quantum gravity near the Planck scale they do appear to be quite plausible and more consistent than the standard results.

Figure captions

Figure 1. Temperature of a black hole versus the mass. Mass is in units of the Planck mass and temperature is in units of the Planck energy. The lower curve is the Hawking result, and the upper curve (with o) is the result using the GUP.

Figure 2. Entropy of a black hole versus the mass. Entropy is dimensionless and mass is in units of the Planck mass. The upper curve is the Hawking result, and the lower curve (with o) is the result using the GUP.

Figure 3. The mass of the black hole versus time. The mass is in units of the Planck mass and the time is in units of the characteristic time. The upper curve is the Hawking result and the lower (with o) is the result using the GUP.

Figure 4. The radiation rate versus time. The rate is in units of the Planck mass per characteristic time. The lower curve is the Hawking result and the upper (with o) is the present result with the GUP.

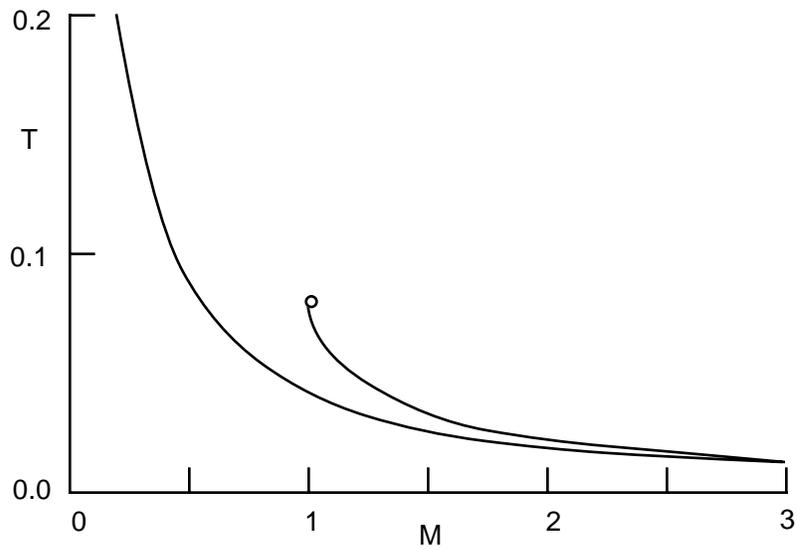

Figure 1. Temperature of a black hole versus the mass. Mass is in units of the Planck mass and temperature is in units of the Planck energy. The lower curve is the Hawking result, and the upper curve (with o) is the result using the GUP.

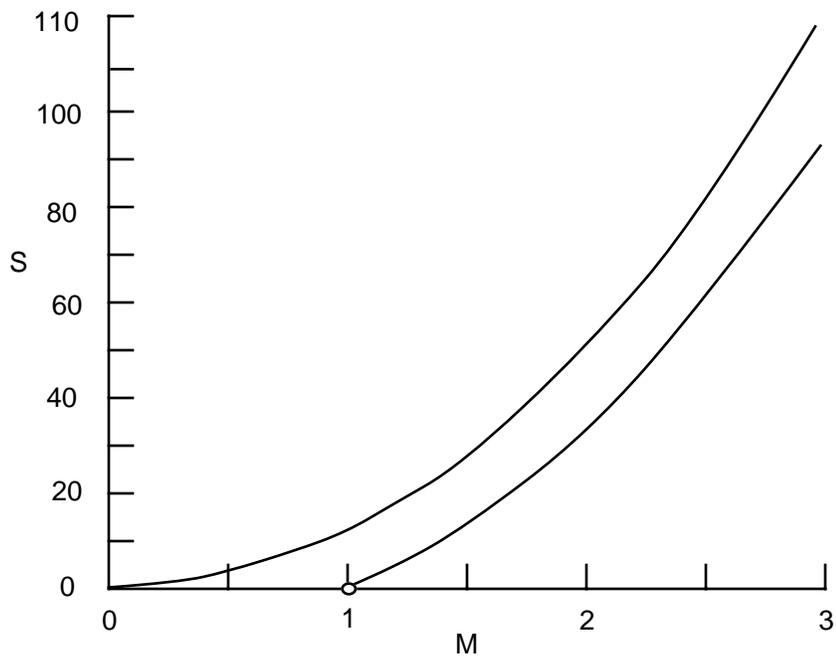

Figure 2. Entropy of a black hole versus the mass. Entropy is dimensionless and mass is in units of the Planck mass. The upper curve is the Hawking result, and the lower curve (with o) is the result using the GUP.

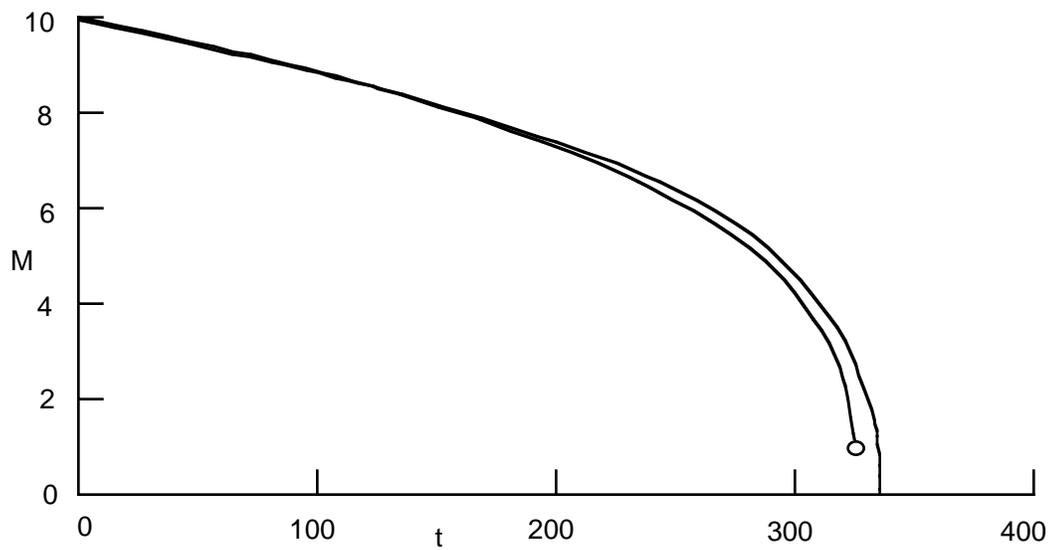

Figure 3. The mass of the black hole versus time. The mass is in units of the Planck mass and the time is in units of the characteristic time. The upper curve is the Hawking result and the lower (with o) is the result using the GUP.

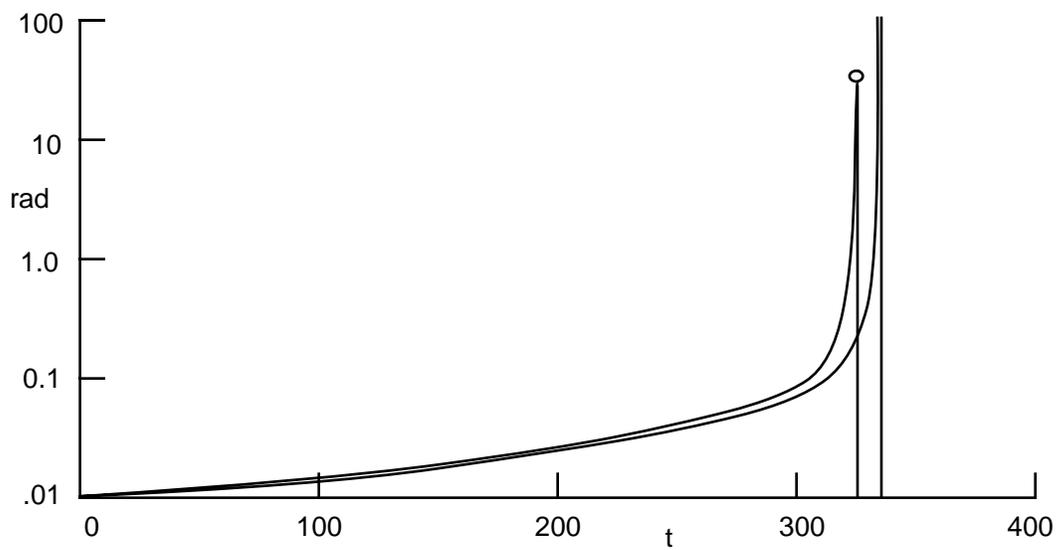

Figure 4. The radiation rate versus time. The rate is in units of the Planck mass per characteristic time. The lower curve is the Hawking result and the upper (with o) is the present result with the GUP.

11